\documentclass[twocolumn,preprintnumbers,amsmath,amssymb]{revtex4}

\usepackage{amssymb}
\usepackage{graphicx}
\usepackage{dcolumn}
\usepackage{amsmath}
\usepackage{subfigure}
\usepackage{bm}
\usepackage[latin1]{inputenc}
\newcommand{\comment}[1]{}

\begin{document}




\title{On the scaling of temperature fluctuations induced by frictional heating}


\author{ Wouter J.T. Bos$^1$,  Robert Chahine$^1$,  Andrey V. Pushkarev$^2$}

\affiliation{$^1$ LMFA, CNRS, Ecole Centrale de Lyon, Universit\'e de Lyon, 
  69134 Ecully, France\\
$^2$Institute of Theoretical and Applied Mechanics, Novosibirsk State University, Novosibirsk, Russia\\
}

\begin{abstract}
The temperature fluctuations generated by viscous dissipation in an isotropic turbulent flow are studied using direct numerical simulation. It is shown that their scaling  with Reynolds number is at odds with predictions from recent investigations.  The origin of the discrepancy is traced back to the anomalous scaling of the dissipation rate fluctuations. Phenomenological arguments are presented which explain the observed results. The study shows that previously proposed models underpredict the variance of frictional temperature fluctuations by a factor proportional to the square of the Taylor-scale Reynolds number. 
\end{abstract}


\maketitle


\section{Introduction}

Two recent investigations have considered the viscous generation of temperature fluctuations in isotropic turbulent flows \cite{Demarinis2013,Bos2014-1}. Even though the models fundamentally differ in the form of the heat production term, they predict the same scaling of the wavenumber spectrum of the heat fluctuations in the case of a statistically steady state. In the second of these works concerns were expressed about the capability of the model to correctly predict the viscous heat production since it was derived using the Direct Interaction Approximation (DIA) \cite{KraichnanDIA}, which is incapable of predicting the cumulant contributions to the wavenumber spectrum of the dissipation rate fluctuations \cite{Chen1989}. Indeed, the temperature fluctuations generated by frictional heating are intimately related to the fluctuations of the dissipation.  In the present work we will first revisit the derivation presented in reference \cite{Bos2014-1} to propose an alternative description of the physics of 
turbulent 
frictional heating. After that, we will check the different assumptions using Direct Numerical Simulations.

\section{Analysis of the problem and scaling arguments}

\subsection{Governing equations}

Using the Reynolds decomposition we write the equation for the fluctuation of the temperature $\theta=\Theta-\bar\Theta$,
\begin{equation}\label{eq:theta}
 \frac{\partial  \theta}{\partial t}+u_i  \frac{\partial\theta}{\partial x_i}=\alpha \frac{\partial^2\theta}{\partial x_i^2}+\frac{1}{c_p}\epsilon',
\end{equation}
where a bar indicates an ensemble average, $c_p$ is the specific heat, $\alpha$ the thermal diffusivity, and $u_i$ an isotropic turbulent velocity field. The fluctuation of the dissipation rate is given by
\begin{equation}\label{eq:eps'}
\epsilon'=\nu\left( \frac{\partial u_i}{\partial x_j}\frac{\partial u_i}{\partial x_j}+\frac{\partial u_i}{\partial x_j}\frac{\partial u_j}{\partial x_i}\right)-
\nu \overline{\frac{\partial u_i}{\partial x_j}\frac{\partial u_i}{\partial x_j}}.
\end{equation}
  For notational convenience $\epsilon$ will be used to denote the average dissipation rate. Since we will not use the total dissipation rate, the absence of a bar on this quantity will not introduce any ambiguity.

The temperature distribution over different lengthscales is given by the temperature spectrum, defined such that
\begin{equation}
 \int E_\theta(k) dk =\overline{\theta^2},
\end{equation}
and the mean dissipation of heat fluctuations is related to this quantity by the relation
\begin{equation}
 \int 2\alpha k^2 E_\theta(k) dk =\epsilon_\theta.
\end{equation}
The evolution-equation of $E_\theta(k)$ is
\begin{eqnarray}\label{eq:DEtdt}
\frac{\partial E_\theta(k)}{\partial t}=T_\theta(k)-D_\theta(k)+P_\theta(k),
\end{eqnarray}
where $T_\theta(k)$ is the interscale transfer term, $D_\theta(k)$ the temperature fluctuation dissipation term and  $P_\theta(k)$ the heat-production term. This last term is given by 
\begin{equation}\label{eq:exprPt}
 P_\theta(k)=\frac{4\pi k^2}{c_p}
\left<
\epsilon'(\bm k)\theta(-\bm k) 
\right>.
\end{equation}
Our interest will be, in particular, in the quantities $\overline{\theta^2}$, $\epsilon_\theta$ and $E_\theta(k)$.

\subsection{The viscous heat production as a function of the dissipation rate fluctuations}

In reference \cite{Bos2014-1}, expression (\ref{eq:exprPt}) was modeled using an approach akin to the Direct Interaction Approximation, expressing  $\left<
\epsilon'(\bm k)\theta(-\bm k) 
\right>$ as a function of velocity triad-interactions only. The concerns expressed in that work motivate us to not express this correlation as a function of the velocity, but to leave explicitly the dissipation rate fluctuations in the expression.

An amount of scalar $\theta$ is advected, diffused and produced. The production part, ignoring the deformation of the scalar blob, can be formally written according to (\ref{eq:theta}), as 
\begin{eqnarray}\label{eq:t1}
\theta(\bm x,t)=\theta(\bm x,t|0)+\frac{1}{c_p}\int_0^t \int g_\theta(\bm x,t|\bm y,s)\epsilon'(\bm y,s)d\bm y ds,
\end{eqnarray}
where the time-integral is taken on a Lagrangian trajectory: $\theta(\bm x,t|0)$ is the value of the scalar fluctuation at time $t=0$ in the fluid particle that passes through $\bm x$ at time $t$. The value of $\theta$ is thus determined by the cumulative heating, proportional to the dissipation rate fluctuation, along a trajectory. In this expression $g_\theta(\bm x,t|\bm y,s)$ is thus the Lagrangian scalar Green's-function. 
The temperature-dissipation correlation, related to the production term in (\ref{eq:exprPt}), is then in a statistically steady state,
\begin{eqnarray}\label{eq:t2}
\left<\epsilon'(\bm x,t)\theta(\bm x+\bm r,t)\right>=\frac{1}{c_p}\int_{0}^t \int \left< g_\theta(\bm x+\bm r,t|\bm y,s)\epsilon'(\bm x,t)\epsilon'(\bm y,s)\right>d\bm y ds.
\end{eqnarray}
We do not attempt here to derive a Lagrangian closure theory for this quantity, but from the above expression we understand that the production is related to the product of dissipation-rate fluctuations, integrated over a Lagrangian correlation time. From equation (\ref{eq:t2}), combined with equation (\ref{eq:exprPt}), we suggest therefore that the production term can be modeled as
\begin{equation}\label{eq:Ptheta1}
P_\theta(k)\sim\frac{\tau(k)}{c_p^2} E_\epsilon(k).
\end{equation}
with
$E_\epsilon(k)$ defined such that
\begin{eqnarray}
\int E_\epsilon(k) dk =\overline{\epsilon'^2}.
\end{eqnarray}
The timescale $\tau(k)$ in expression (\ref{eq:Ptheta1}) represents the correlation time of the scalar fluctuation over a Lagrangian trajectory, combined with the correlation time of the dissipation rate fluctuations. It is at present not entirely clear what the functional form of this timescale is. The timescale of inertial range eddies of scale $l\sim 1/k$, consistent with Kolmogorov-Obukhov scaling, is $\tau(k)\sim \epsilon^{-1/3}k^{-2/3}$, which is the Lagrangian timescale that is most commonly used in turbulence theory \cite{Monin}. However, for several phenomena such as Burger's turbulence or the advection of a scalar in the presence of a mean scalar gradient, the integral timescale $\mathcal T\sim \epsilon^{-1/3}L^{2/3}$ ($L$ is the integal lengthscale) can play a dominant role \cite{Kraichnan1968-2,Bos2014-2}. In the present case, as we will see, the dissipation-rate fluctuations are correlated at large scales, so that it is not excluded that the dynamics are governed by $\mathcal T$. In order to leave open the possibility of both timescales to play a role, we introduce the hybrid timescale
\begin{equation}\label{eq:tauH}
 \tau(k)\sim \epsilon^{-1/3}k^{-2/3}(kL)^{\alpha}.
\end{equation}
The parameter $\alpha$ allows to consider the different possibilities. For $\alpha=0$ we have the Kolmogorov-Obukhov timescale and for $\alpha=2/3$ we find the integral timescale.

%
%
%
%
%

Expression (\ref{eq:Ptheta1}) shows that, according to our arguments, the viscous heat-production is proportional to the spectrum of the dissipation rate fluctuations. It is thus this latter quantity that plays a major role in the determination of the viscous heating.

The spectrum $E_\epsilon(k)$ has received considerable interest in the past, in particular since its behavior is largely in disagreement with dimensional scaling within the Kolmogorov framework \cite{Kolmogorov}. In particular, the latter scaling theory would predict the dissipation rate spectrum to behave as 
\begin{equation}\label{eq:Eeps1}
 E^{(1)}_\epsilon(k)\sim \nu^2 \epsilon^{4/3}k^{5/3}.
\end{equation}
This scaling can be obtained by assuming the velocity to satisfy a multi-variate Gaussian distribution and assuming the energy spectrum to have an inertial range proportional to 
\begin{equation}\label{eq:K41}
 E(k)\sim \epsilon^{2/3}k^{-5/3}.
\end{equation}
Experiments \cite{Gurvich1963,Pond1965,Atta1970} (see \cite{Monin} for a review) and simulations \cite{Gotoh1999} show howevever that the scaling (\ref{eq:Eeps1}) is not observed at high Reynolds numbers and that, instead of an increasing function of the wavenumber, $E_\epsilon(k)$ is a decreasing function of $k$. Yaglom proposed a model which fits the data qualitatively, taking into account the non-Gaussian character of the dissipation rate \cite{Yaglom1966,Monin}. His model predicts the spectrum
\begin{equation}\label{eq:Eeps2}
 E^{(2)}_\epsilon(k)\sim \epsilon^{2}L(kL)^{-1+\mu},
\end{equation}
where $L$ is a large-scale correlation length, and $\mu$ is an intermittency parameter. In reference [\onlinecite{Monin}] values are reported in the interval $0.3<\mu<0.5$. Since then, a number of different models have been proposed to more precisely predict the physics of the dissipation rate, but these refinements are beyond the scope of the present investigation, given the accuracy obtained in the determination of scaling exponents at low and moderate Reynolds numbers. 

Substituting expression (\ref{eq:Eeps1}) in the expression for the heat production term (\ref{eq:Ptheta1}) and using $\tau(k)\sim \epsilon^{-1/3}k^{-2/3}$ we find for the Gaussian estimate of the production term,
\begin{equation}\label{eq:Ptheta100}
P^{(1)}_\theta(k)\sim\frac{\nu^2\epsilon k}{c_p^2},
\end{equation}
and this is what was obtained in reference \cite{Bos2014-1}.

If we use however the scaling proposed by Yaglom (\ref{eq:Eeps2}), combined with the timescale (\ref{eq:tauH}), we find the production term proportional to
\begin{equation}\label{eq:Ptheta200}
P^{(2)}_\theta(k)\sim\frac{\epsilon^{5/3}k^{-5/3+\mu+\alpha}L^{\mu+\alpha}}{c_p^2},
\end{equation}
and we recall that that $\mu$ is the intermittency parameter in the range $0.3<\mu<0.5$, and $\alpha$ characterizes the timescale ($\alpha=0$ for a Kolmogorov-Obukhov timescale, $\alpha=2/3$ for the integral timescale).

While getting heated on its trajectory, the fluid blob will also be deformed, thereby exchanging its temperature variance with other lengthscales. This effect is represented by the term $T_\theta(k)$ in expression (\ref{eq:DEtdt}).  We will investigate the scaling of the temperature spectrum, as in \cite{Bos2014-1}, by assuming the scalar transfer to be given by a Kovaznay-type scalar transfer model \cite{Rubinstein2013}, which allows simple analytical treatment. This model reads
\begin{equation}\label{eq:Kovaz}
 T_\theta(k)\sim \frac{\partial}{\partial k}\left(E_\theta(k)E(k)^{1/2}k^{5/2}\right).
\end{equation}
In the statistically stationary case, in the range where the production is important, a balance is expected between the transfer  and the production,
\begin{equation}\label{eq:TisP}
T_\theta(k)\approx P_\theta(k). 
\end{equation}
We will consider the case of unity Prandtl number, $\nu=\alpha$, and we suppose the spectra to extend from $k_L\sim 1/L$ to $k_\eta$, with $k_\eta\sim \epsilon^{1/4}\nu^{-3/4}$ and $k_\eta\gg k_L$, and to display powerlaw behaviour throughout. The energy spectrum is given by (\ref{eq:K41}) and the production spectrum is given either by (\ref{eq:Ptheta100}) or (\ref{eq:Ptheta200}). 
Integrating (\ref{eq:TisP})  from $k_L$ to $k$, we have from (\ref{eq:Kovaz}) and (\ref{eq:K41}) 
\begin{equation}
 E_\theta(k)\sim \epsilon^{-1/3}k^{-5/3}\int_{k_L}^k P_\theta(k)dk.
\end{equation}
For the Gaussian estimate of the production spectrum (\ref{eq:Ptheta100}) and the Kolmogorov-Obukhov time $\tau(k)\sim \epsilon^{-1/3}k^{-2/3}$ this gives for $k\gg k_L$,
\begin{equation}\label{eq:ET1}
E^{(1)}_\theta(k)\sim \frac{\epsilon^{2/3}\nu^{2}k^{1/3}}{c_p^2}.
\end{equation}
For the production term taking into account intermittency effects (\ref{eq:Ptheta200}) we find 
\begin{equation}\label{eq:ET2}
E^{(2)}_\theta(k)\sim \frac{\epsilon^{4/3}L^{2/3}k^{-5/3}}{c_p^2}\left((kL)^{\mu+\alpha-2/3}-1\right).
\end{equation}
We now consider the different values of the exponent for the timescale $\tau$. We find for $kL\gg 1$
\begin{equation}\label{eq:ET2,0}
E^{(2)}_\theta(k)\sim \frac{\epsilon^{4/3}L^{2/3}k^{-5/3}}{c_p^2}~~\textrm{for $\alpha=0$},
\end{equation}
independent of $\mu$. Alternatively, it gives
\begin{equation}\label{eq:ET2,2/3}
E^{(2)}_\theta(k)\sim \frac{\epsilon^{4/3}L^{2/3}k^{-5/3}}{c_p^2}(kL)^{\mu}~~\textrm{for $\alpha=2/3$},
\end{equation}
and the spectrum becomes thus proportional to $k^{-5/3+\mu}$. The fact that, depending on the value of $\alpha$, expression (\ref{eq:ET2}) yields two qualitatively different spectra (one dependent on $\mu$, the other not), stems from the fact that the forcing term (\ref{eq:Ptheta200}) is a multiscale (or fractal) forcing, with a powerlaw exponent smaller than $-1$ for $\alpha=0$, but larger than $-1$ for $\alpha=2/3$. In the former case the spectral flux, determined by the integral of the production up to wavenumber $k$, becomes then independent of the forcing exponent. For the above integrals to converge we need to satisfy the constraint $0<\mu<2/3$, which includes the experimental interval $0.3<\mu<0.5$.

%
%

The variance of the temperature fluctuations can be computed by integrating the spectra $E^{(1)}_\theta(k)$ and $E^{(2)}_\theta(k)$ between $k_L$ and $k_\eta$. 
For the two different predictions (\ref{eq:ET1}) and (\ref{eq:ET2}) we find,
\begin{eqnarray}
 \overline{\theta^2}^{(1)}\sim \frac{\epsilon \nu}{c_p^2} ,\label{eq:t2eps}\\
 \overline{\theta^2}^{(2)}\sim \frac{(\epsilon L)^{4/3}}{c_p^2},\label{eq:t2eps2}
\end{eqnarray}
and both values $\alpha=0,2/3$ give the same scaling for the variance of the temperature fluctuations $\overline{\theta^2}^{(2)}$, independent of the value of $\mu$. This is not so for the destruction of scalar variance,
\begin{equation}
  \epsilon_\theta=\int P_\theta(k)dk.
\end{equation}
For the Gaussian case it gives 
\begin{equation}
\epsilon_\theta^{(1)}\sim \frac{\epsilon^{3/2} \nu^{1/2}}{c_p^2}. 
\end{equation}
For the non-Gaussian estimate it yields
\begin{eqnarray}
 \epsilon_\theta^{(2)}&\sim \frac{\epsilon^{5/3} L^{2/3}}{c_p^2} ~~&\textrm{for $\alpha=0$}\\
                      &\sim \frac{\epsilon^{5/3} L^{2/3}}{c_p^2}R_L^{3\mu/4} ~~&\textrm{for $\alpha=2/3$},
\end{eqnarray}
where $R_L=UL/\nu$ and where we used $\epsilon\sim U^3/L$ with $U$ the RMS velocity. The normalized variance and scalar-dissipation can then be defined by 
\begin{eqnarray}
\tilde{{\theta^2}}^{(1)}\sim \overline{\theta^2}\frac{ c_p^2}{\epsilon \nu},~~~~~~  \tilde{\epsilon}_\theta^{(1)}\sim \epsilon_\theta\frac{ c_p^2}{\epsilon^{3/2} \nu^{1/2}},\label{eq:t2epstilde}\\
 \tilde{{\theta^2}}^{(2)}\sim \overline{\theta^2}\frac{c_p^2}{(\epsilon L)^{4/3}},~~~~~~   \tilde{\epsilon}_\theta^{(2)}\sim \epsilon_\theta \frac{c_p^2}{\epsilon^{5/3} L^{2/3}} \label{eq:t2eps2tilde}. 
\end{eqnarray}
The quantities  $\tilde{{\theta^2}}^{(1)}$, $\tilde{{\theta^2}}^{(2)}$ and $\tilde{\epsilon}_\theta^{(1)}$ should become constant at large values of the Reynolds number, if the underlying assumptions are correct.  The quantity $\tilde{\epsilon}_\theta^{(2)}$ should become constant if $\alpha=0$. However, if $\alpha=2/3$, it should display a Reynolds number dependence, proportional to $R_L^{3\mu/4}$.

Both the wavenumber scaling of the temperature spectrum and the Reynolds number dependence of the integrated quantities depend thus strongly on the scaling of $E_\epsilon(k)$. The scaling obtained for $E^{(1)}_\theta(k)$, $\overline{\theta^2}^{(1)}$ and $\epsilon_\theta^{(1)}$ is identical to the one observed in closure simulations \cite{Bos2014-1}. In the following it will  be investigated by DNS which scaling is observed when the discretized Navier-Stokes equations are numerically solved. To evaluate the Reynolds number dependence, we will fix the forcing lengthscale $L$ and vary the viscosity.

\section{Simulations of statistically stationary isotropic turbulence with frictional heat production}
 
\subsection{Numerical method and required accuracy}

We solve the Navier-Stokes equations with a large-scale forcing term,
\begin{equation}
 \frac{\partial u_i}{\partial t}+u_j \frac{\partial u_i}{\partial x_j}=-\frac{1}{\rho}\frac{\partial p}{\partial x_i}+\nu \frac{\partial^2 u_i}{\partial x_j^2}+f_i,
\end{equation}
and the equation of the total (average plus fluctuating) temperature,
\begin{equation}\label{eq:Theta}
 \frac{\partial  \Theta}{\partial t}+u_i  \frac{\partial\Theta}{\partial x_i}=\alpha \frac{\partial^2\Theta}{\partial x_i^2}+\frac{\nu}{c_p}\left( \frac{\partial u_i}{\partial x_j}\frac{\partial u_i}{\partial x_j}+\frac{\partial u_i}{\partial x_j}\frac{\partial u_j}{\partial x_i}\right),
\end{equation}
in a cubic three-dimensional periodic domain of size $2\pi$, using a standard pseudo-spectral solver. The forcing is introduced by a negative viscosity, acting on the modes with wavenumbers smaller than $2.5$. The initial temperature field is zero and the initial velocity field consists of random noise. All results are evaluated once a statistically steady state is reached were the velocity and temperature fluctuate with a constant variance, and $\bar\Theta\sim t$. Once this state is reached, results for $\overline{\theta^2}$, $\epsilon_\theta$, $\overline{u_i^2}$ and $\epsilon$ are obtained by averaging over a sufficiently large time-interval to have the errors induced by the finiteness of the averaging time to be smaller than errors induced by the discretization of the domain. In order to perform this time-averaging over a long enough time-interval (70 integral timescales $\mathcal T=e_{kin}/\epsilon$, with $e_{kin}$ the kinetic energy), the simulations were carried out at a moderate maximum resolution of 
$256^3$ grid-points.

It was observed that the results were very sensitive to the resolution of the simulations. This issue is well known for DNS studies of dissipation-range quantities \cite{Watanabe2007-2,Donzis2008}. Whereas the kinetic energy and dissipation were reasonably well resolved for simulations with $k_\textrm{max}\eta\approx 1$, the temperature statistics needed about twice this resolution. Simulations were therefore carried out using $k_\textrm{max}\eta\approx 2$. The discretization errors were estimated by comparing the measured quantities at the highest resolutions at $k_\textrm{max}\eta\approx 2$ and $1.5$, respectively,  $\varepsilon_f=|f_2-f_{1.5}|/|f_2|$ for a given Reynolds number $R_\lambda=61$. The errors in the kinetic energy, dissipation rate, scalar variance and scalar dissipation were hereby estimated to be $1\%$, $2\%$, $14\%$, $9\%$ respectively. These errors will be indicated by error-bars in Figure \ref{fig:var}. Even though these relative errors are quite large, the scaling behaviour of the 
wavenumber-spectra was not significantly affected. 

\subsection{Results on the scaling of the velocity and dissipation rate spectra}

\begin{figure}[]
\center{
\includegraphics[width=.5\textwidth,angle=0]{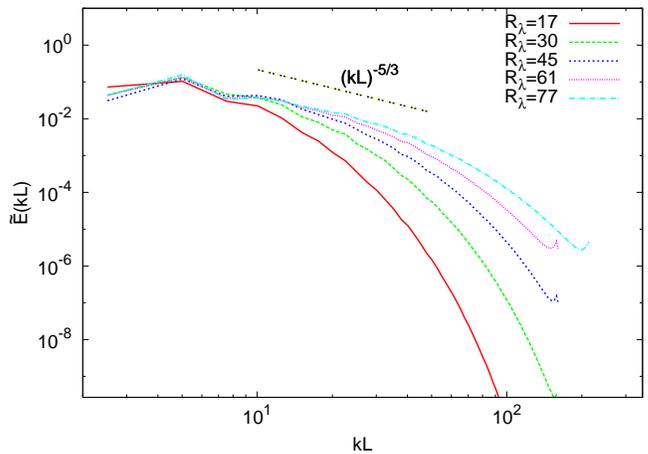}}
\vspace{0.7cm}
\caption{Energy spectra at different Reynolds numbers, normalized using the  quantities $\epsilon,L$. 
}
\label{Fig:Ek}
\end{figure}

As was first pointed out by Taylor \cite{Taylor1935}, the dissipation rate becomes independent of the viscosity at high Reynolds number. The dissipation rate depends then only on the large scale quantities $L$ and $U$ where $U$ is the root-mean-square value of a velocity component. The relation is 
\begin{equation}\label{eq:0th}
 \epsilon\sim \frac{U^3}{L},
\end{equation}
where the proportionality constant is independent of the Reynolds number but depends on the large scale flow \cite{Bos2007-2}. Since the dissipation rate depends only on the large scales, the energy spectrum (\ref{eq:K41}), normalized by the quantities $\epsilon$ and $L$, 
\begin{equation}
\tilde E(kL)\equiv \frac{E(k)}{\epsilon^{2/3} L^{5/3}} \sim (kL)^{-5/3}.
\end{equation}
should therefore collapse at high Reynolds numbers in the forcing- and the inertial-range, if the large-scale forcing mechanism is not a function of the Reynolds number. This is shown in Figure \ref{Fig:Ek} to be the case already at the relatively low values of the Reynolds number considered in the present work. We note that often in literature the scaling of spectral energy distributions is displayed in Kolmogorov-variables, i.e., length- and timescales based on $\epsilon$ and $\nu$. This allows a collapse of the spectra in the dissipation range. However, in the present case, the use of the Tayor-variables $L$ and $U$ to collapse the normalized spectra in the forcing scales is more efficient to decide which scaling describes better the spectra of the dissipation rate and heat fluctuations.


\subfigcapskip=20pt
\begin{figure}[]
\center{
\subfigure[]{\includegraphics[width=.5\textwidth,angle=0]{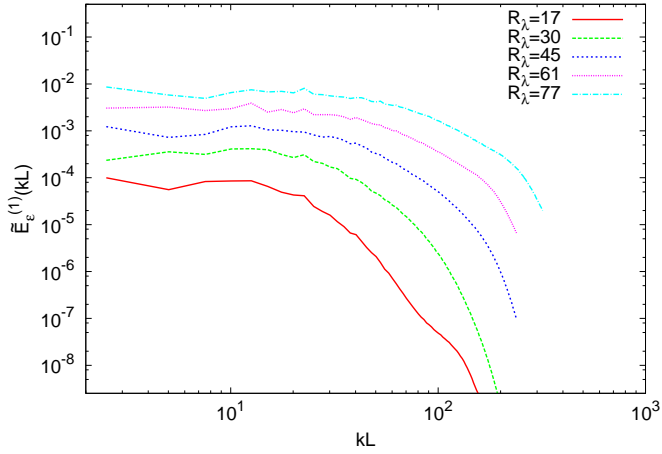}}
\subfigure[]{\includegraphics[width=.5\textwidth,angle=0]{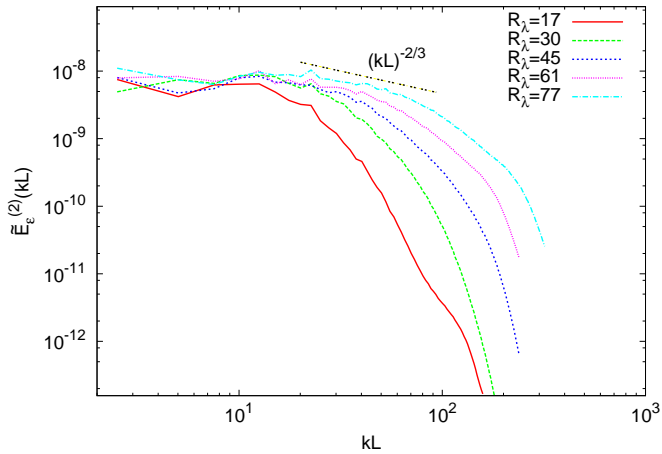}}
}
\caption{(a) Dissipation fluctuation spectra at different Reynolds numbers, normalized using the quantities $\epsilon,L,\nu$, assuming expression (\ref{eq:Eeps1}) to hold. (b) The same spectra normalized according to expression (\ref{eq:Eeps2}).}
\label{Fig:EepsL}
\end{figure}

The spectrum of the dissipation rate fluctuations is, according to expression (\ref{eq:Eeps1}), proportional to $\nu^2$. If the large scales of the velocity field are independent of the Reynolds number, as suggested by the results in Figure  \ref{Fig:Ek} (a), this would imply that the normalized spectra would scale as
\begin{equation}
\tilde E^{(1)}_\epsilon(kL)\equiv\frac{E_\epsilon(k)L^{5/3}}{\nu^{2}\epsilon^{4/3}}\sim (kL)^{5/3}
\end{equation}
and collapse in the large scales. It is shown in Figure \ref{Fig:EepsL} (a) that this is not the case. According to expression (\ref{eq:Eeps2}), the spectra should rather be normalized as
\begin{equation}
\tilde E^{(2)}_\epsilon(kL)\equiv\frac{E_\epsilon(k)}{\epsilon^{2}L}\sim (kL)^{-1+\mu}.
\end{equation}
It is shown in Figure \ref{Fig:EepsL} (b) that this gives a good collapse of the data. Also, the value of $\mu$ is of order $1/3$, but the precise value is hard to determine at these values of the Reynolds number. From these results we can conclude that the model proposed by Yaglom is sufficiently accurate to roughly describe the Reynolds number dependency of the dissipation-rate fluctuation spectrum observed in the present work.  
 
\subsection{Results on the scaling of the temperature fluctuations}

\begin{figure}[]
\center{
\subfigure[]{\includegraphics[width=.5\textwidth,angle=0]{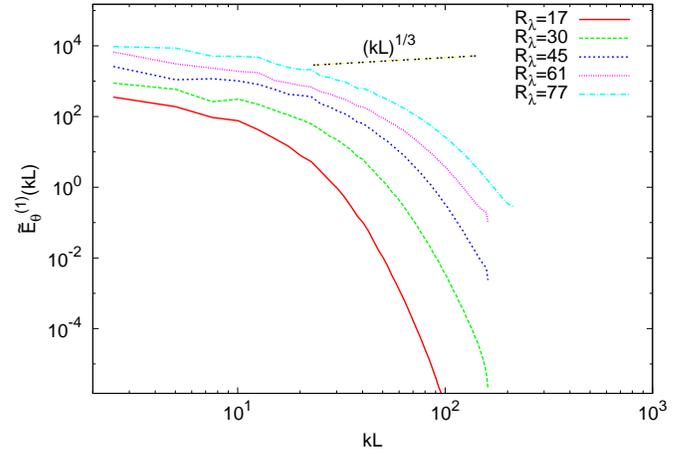}}
\subfigure[]{\includegraphics[width=.5\textwidth,angle=0]{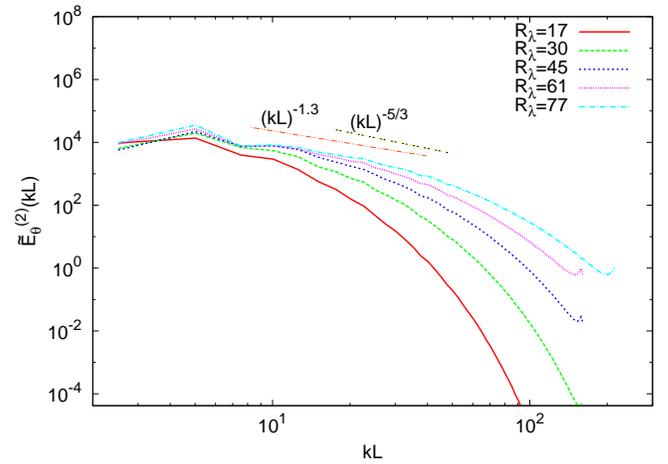}}
}
\caption{Temperature spectrum for different Reynolds numbers. (a) Normalized using expression (\ref{eq:Et1normL})). (b) Normalized using the expression (\ref{eq:Et2normL}).}
\label{Fig:ETL}
\end{figure}

In order to verify which of the predictions (\ref{eq:ET1}) or (\ref{eq:ET2}) is correct, we use the same arguments as used in the previous section for the spectrum $E_\epsilon(k)$. Indeed, normalizing the scaling (\ref{eq:ET1}) we should have
\begin{equation}\label{eq:Et1normL}
\tilde E^{(1)}_\theta(kL)\equiv \frac{c_p^2 L^{1/3}}{\epsilon^{2/3}\nu^2} E^{(1)}_\theta(k)\sim (kL)^{1/3}.
\end{equation}
Using scaling (\ref{eq:ET2}), we have 
\begin{eqnarray}
\tilde E^{(2)}_\theta(kL)\equiv\frac{c_p^2}{\epsilon^{4/3}L^{7/3}}E^{(2)}_\theta(k)\sim (kL)^{-5/3}~~\textrm{for $\alpha=0$},\label{eq:Et2normL}\\
\sim (kL)^{-5/3+\mu}~~\textrm{for $\alpha=2/3$}\label{eq:Et2normL2}.
\end{eqnarray}
Clearly, comparing Figures \ref{Fig:ETL} (a) and (b) we see that the results are in far better agreement when using the second scaling relation. In particular the collapse at small values of $kL$ is significantly better using scaling (\ref{eq:Et2normL}) or (\ref{eq:Et2normL2}) than for scaling (\ref{eq:Et1normL}). The question whether the inertial range timescale is a function of the wavenumber or not is not easily answered from this representation. Indeed the spectrum does not seem incompatible with a scaling proportional to $(kL)^{-5/3+\mu}$ (we have added a line with a powerlaw exponent of $-1.3$), but the inertial range is not sufficiently large to give a conclusive answer.
 
\begin{figure}[]
\center{
\subfigure[]{\includegraphics[width=.5\textwidth,angle=0]{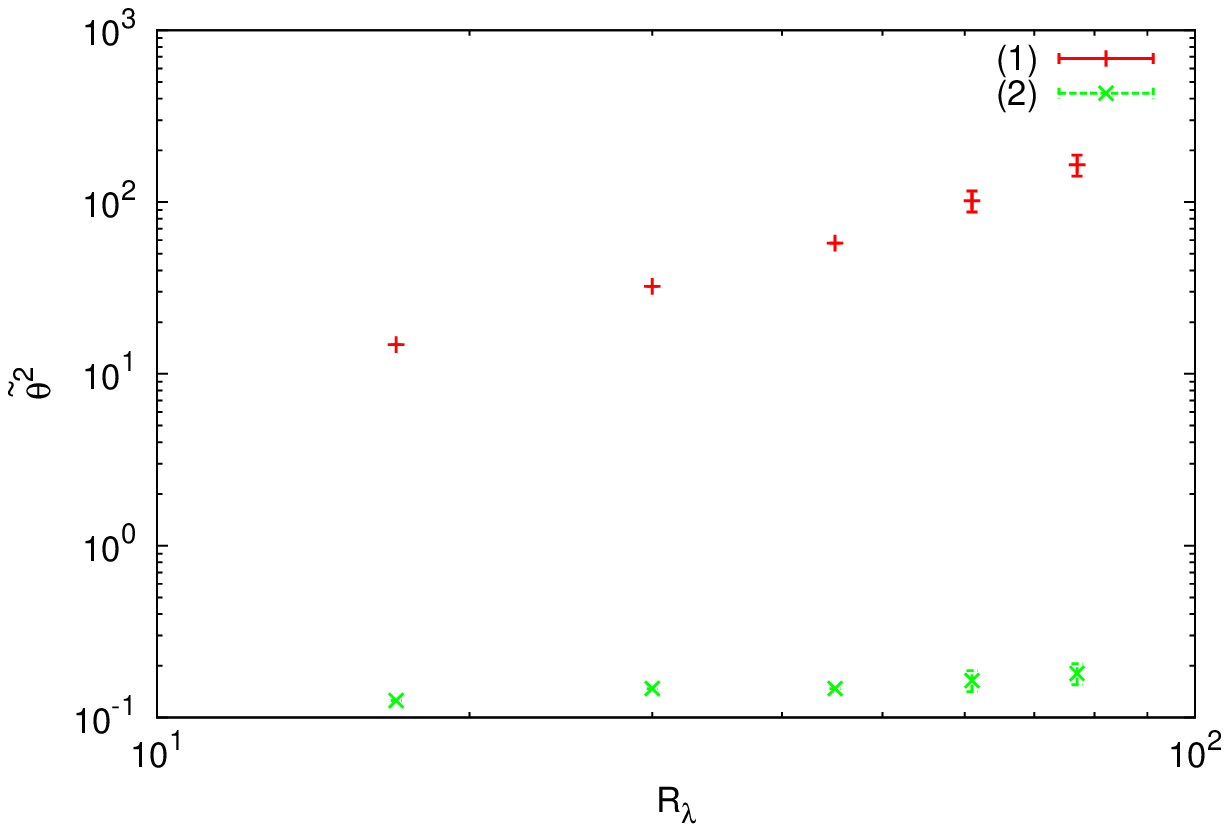}}
\subfigure[]{\includegraphics[width=.5\textwidth,angle=0]{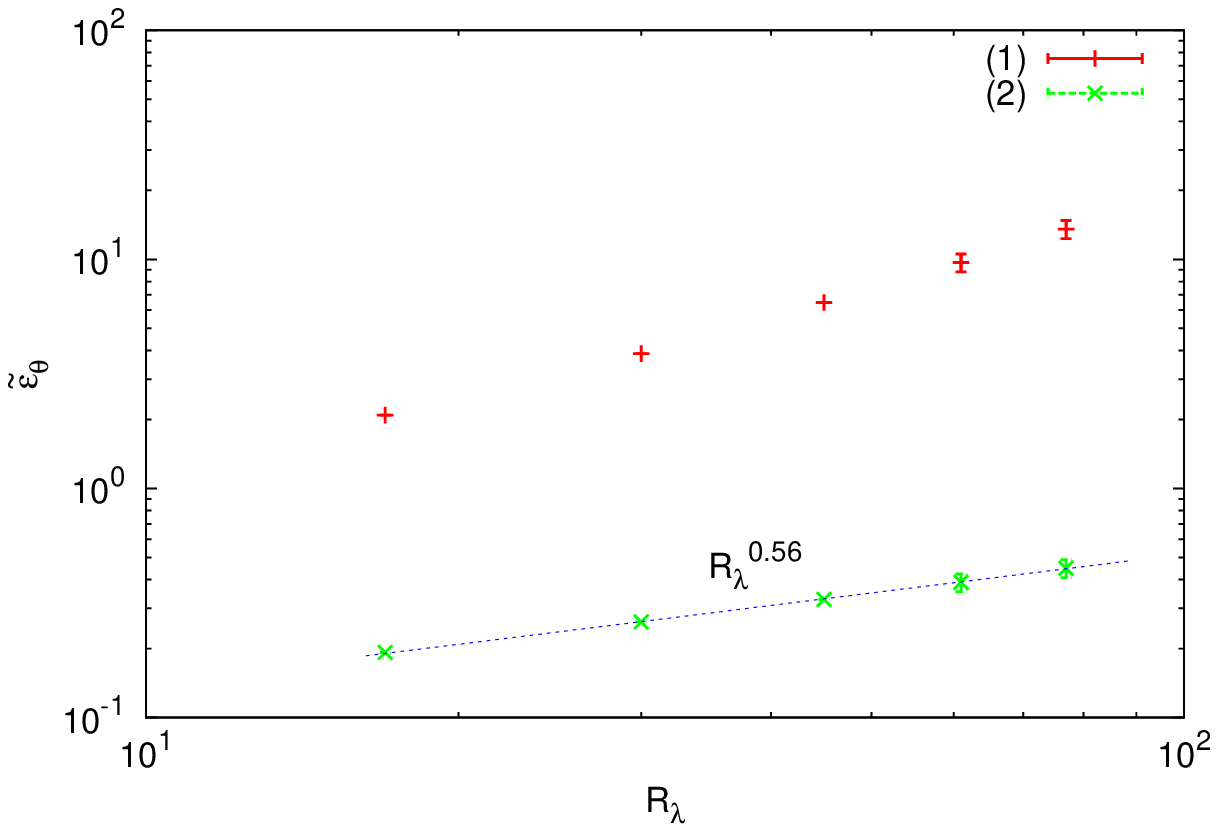}}
}
\caption{Reynolds number dependence of the normalized temperature variance and the normalized dissipation of temperature fluctuations. Normalizations are given in expressions (\ref{eq:t2eps}) and (\ref{eq:t2eps2}).}
\label{fig:var}
\end{figure}

\begin{figure}[]
\center{
\subfigure[]{\includegraphics[width=.4\textwidth,angle=0]{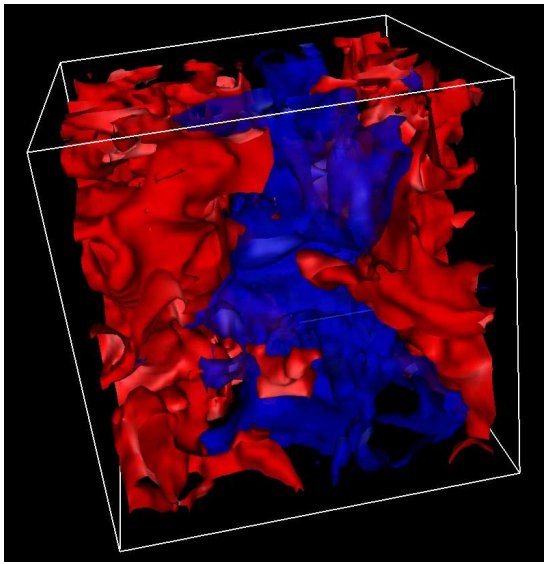}}
\subfigure[]{\includegraphics[width=.4\textwidth,angle=0]{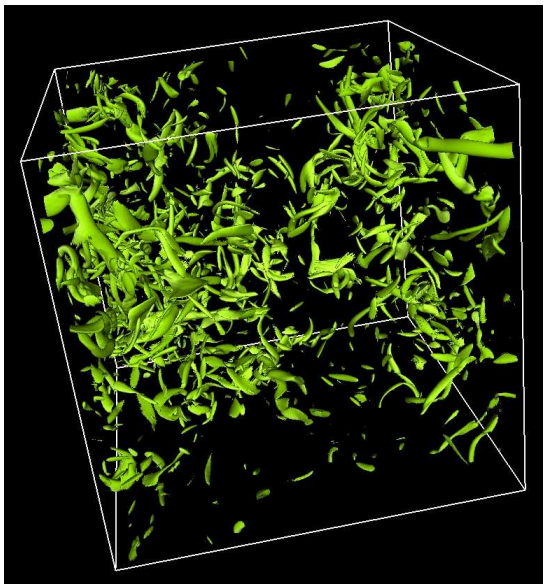}}
}
\caption{Visualizations of (a) the temperature field (red isosurfaces correspond to positive heat fluctuations: $\theta/\theta_{rms}=0.83$; the blue ones to negative fluctuations $\theta/\theta_{rms}=-1.26$.) (b) the vorticity field (iso-surfaces of the enstrophy for $\|\bm \omega \|/\omega_{rms}=2.77$). The Reynolds number is $R_\lambda=77$. Visualizations by VAPOR \cite{Clyne2007}.}
\label{Fig:visu}
\end{figure}

In Figure \ref{fig:var} we show the Reynolds number dependence of $\tilde{{\theta^2}}$ and $\tilde\epsilon_\theta$, using the normalizations proposed in expressions (\ref{eq:t2epstilde}) and (\ref{eq:t2eps2tilde}). The normalized variances should, according to our predictions, become independent of the Reynolds number. This is very clearly the case for $\tilde{{\theta^2}}^{(2)}$, which confirms the results observed for the normalized spectra.  We recall that this quantity is not sensitive to the shape of the timescale. For the dissipation of the temperature fluctuations, the normalization $\tilde{\epsilon}_\theta^{(2)}$ is closer to a constant value than $\tilde{\epsilon}_\theta^{(1)}$, but a clear Reynolds number dependence is observed. We recall that according to the prediction using the integral timescale ($\alpha=2/3$) in the expression for the heat-fluctuation production-term, the normalized temperature-dissipation should scale as $R_\lambda^{3\mu/2}$. A best fit of a powerlaw through the data-points in Figure \ref{fig:var}~b gives a dependence of $R_\lambda^{0.56}$, corresponding to a value of $\mu=0.37$. Note that this value of $\mu$ gives a scalar spectrum proportional to $k^{-1.3}$, as indicated in Figure \ref{Fig:ETL}~(b). It seems that the non-Gaussian model combined with the integral timescale for the Lagrangian correlation time $\tau(k)$ is capable of explaining all the numerical results.

\section{Discussion}

The intermittent character of the dissipation rate fluctuations affects dramatically the spectrum of the temperature fluctuations. The wavenumber dependence of the temperature spectrum is shown to be strongly correlated at large scales, whereas the Gaussian estimate and closure expressions of the Eddy-Damped Quasi-Normal Markovian type mispredict this spectrum to have a correlation-length of the order of the dissipation scale.  Our results further indicate that the timescale governing the dynamics of the heat-production is the integral timescale and not a timescale of the form $\epsilon^{-1/3}k^{-2/3}$. The intermittency coefficient appearing in Yaglom's model for the dissipation-rate fluctuations is of the order $\mu\approx 0.3$, if it is estimated from the Reynolds number dependence of the temperature dissipation.

One could speculate on the underlying physical processes in terms of flow structures to explain the observed results. Indeed, vortices are elongated structures which have a diameter roughly between the dissipation lengthscale and the Taylor lengthscale, depending on the precise definition, and a length which can extend up to to the integral lengthscale. It is tempting to relate this last lengthscale to the correlation-length of the temperature fluctuations, arguing that they are produced in the strong velocity gradient regions around these vortices. 

Another phenomenological explanation could start from the 
observation that clusters of vorticity are correlated over the integral lengthscale, a picture which was used to model the intermittent character of the pressure fluctuations and dissipation-rate fluctuations in \cite{Gotoh1999}. In order to rigorously investigate such mechanisms, precise criteria should be introduced to define structures and clusters, and this is beyond the scope of the present work. However, as a first step, we present in Figure \ref{Fig:visu} flow visualizations, where we show iso-vorticity surfaces and surfaces of temperature fluctuations. The main observation is that, in agreement with the spectral distribution of temperature fluctuations, the volume rendering shows large-scale correlated smooth temperature iso-surfaces. The enstrophy fluctuations show their typical worm-like small-scale structure. Comparing figures (a) and (b), a correlation seems to exist between regions of intense vorticity clustering and positive temperature fluctuations, but more quantitative measures are needed to 
assess this.   

With respect to the predictions of closure \cite{Bos2014-1}, we can estimate what the difference in realistic flows will be for the temperature variance. Combining expression (\ref{eq:0th}) with (\ref{eq:t2eps}) and (\ref{eq:t2eps2}) we find that the variance according to the two different model predictions scales as
\begin{eqnarray}
 \overline{\theta^2}^{(1)}\sim \frac{U^4}{c_p^2} R_\lambda^{-2}\nonumber\\
 \overline{\theta^2}^{(2)}\sim  \frac{U^4}{c_p^2} R_\lambda^{0}.
\end{eqnarray}
If we consider a perfectly insulated flow, the mean temperature will, in a steady state, increase linearly in time, since
\begin{equation}
\frac{\partial \bar \Theta}{\partial t} =\frac{\epsilon}{c_p}.
\end{equation}
The timescale over which the temperature increases, normalized by the temperature variance is then, according to (\ref{eq:t2eps2}),
\begin{equation}\label{eq:dTdt}
\frac{\sqrt{\overline{\theta^2}}}{\partial \bar\Theta/\partial t} \sim L^{2/3}\epsilon^{-1/3},
\end{equation}
which corresponds to the integral timescale. This means that the rms fluctuations will be of the order of the mean temperature-increase over one integral timescale.  In the closure prediction, the timescale in (\ref{eq:dTdt}) was of the order of the Kolmogorov-scale, so that for a moderately turbulent flow, the rms fluctuations were expected to be a factor $R_\lambda$ smaller in a practical situation and measurement of the temperature fluctuations seemed out of reach in a turbulent flow for that case. According to the predictions of the present work, such measurements will be possible, though challenging. 

A question with respect to turbulence theory is how these effects could be captured by statistical theory derived from the Navier-Stokes equations. In particular, which assumptions should be modified in the Direct Interaction Approximation to capture such effects. Progress on this seems to have been small since the later works of Kraichnan \cite{Kraichnan1989}.

\section*{Acknowledgments}

Interaction with Robert Rubinstein and insightful suggestions of the referees are gratefully acknowledged. We further acknowledge support from the French National Research Agency through project SiCoMHD. CPU time was provided by the PCMS2I.


\end{document}